\documentclass[preprint,showpacs,aps]{revtex4}

\usepackage{epsfig}
\usepackage{color}

\begin{document}
\title{\large \bf Dynamic approach to finite-temperature magnetic
phase transitions in the extended $J_1$-$J_2$ model with a vacancy
order}

\author{N.J. Zhou$^{1}$\footnote{corresponding author; email: zhounengji@hznu.edu.cn}, B. Zheng$^{2}$, and J.H. Dai$^{1}$ }

\affiliation{$^{1}$Department of Physics, Hangzhou Normal University, Hangzhou 310036, P.R. China \\
$^{2} $Department of Physics, Zhejiang University, Hangzhou 310027, P.R. China}

\begin{abstract}
The recently discovered iron-based superconductors
A$_{y}$Fe$_{2-x}$Se$_{2}$ ($A$=K, Rb, Cs, Tl) show a long-range
antiferromagnetic order with an unexpected high transition
temperature $T_N \sim 550$ K and a unique $\sqrt{5} \times \sqrt{5}$
vacancy order.  Taking the extended $J_1$-$J_2$ model as
a minimal model, we investigate the finite-temperature magnetic
phase transitions in a square lattice with a $\sqrt{5} \times
\sqrt{5}$ vacancy superstructure by using large-scale Monte Carlo
simulations. By the parallel tempering technique, the block spin
checkerboard and stripe antiferromagnetic states are detected to be
the groundstates for three representative sets of
model parameters. The short-time dynamic approach is
applied to accurately determine the critical temperature as well as
the static and dynamic exponents. Our results indicate that the
dramatic enhancement of the critical temperature as observed in
experiments should be mainly due to a combination effect of the
vacancy order and the block lattice contraction.

\end{abstract}


\maketitle
\section{Introduction}
The discovery of superconductivity in iron pnictides
\cite{kam08,che08,che08b,ren08b,wan08} has renewed an intensive
study on the interplay between superconductivity and
antiferromagnetism \cite{pag10}. A broad family of the iron-based
superconductors has been synthesized, of which the parent compounds
are typically represented by the $1111$-type LaFeAsO \cite{kam08},
the $122$-type BaFe$_2$As$_2$ \cite{rot08}, the $111$-type LiFeAs
\cite{wan08b}, and the $11$-type FeSe \cite{hsu08}.
Antiferromagnetic transitions occur around the N\'eel
temperature $T_N \approx 100-200$ K, and the long-range magnetic order
is a stripe-like (or collinear) antiferromagnetic state
\cite{cru08,che09} except for the $11$-type where the magnetic order
is a bi-collinear antiferromagnetic state \cite{bao09}. The magnetic
properties as well as superconductivity are closely related to a
common two-dimensional ($2D$) Fe-atom square lattice
\cite{sin09,joh10,maz11}. The effective magnetic moments of each
irons determined in experiments are usually within
$0.3\mu_B$/Fe$-1.0\mu_B$/Fe, while the iron moments estimated from
the first-principles calculations \cite{sin08,cao08,yil08,ma08} or
model analysis \cite{si08} could be around $2.0\mu_B$/Fe or larger.

Recently, a new family of iron-based superconductors, i.e., the
intercalated iron chalcogenides A$_{y}$Fe$_{2-x}$Se$_{2}$ ($A$=K,
Rb, Cs, Tl), has been found with a moderate high
superconducting transition temperature $T_{sc}\sim 30$ K
\cite{guo10,fan11,wan11}. These materials are structurally similar
to the $122$-type iron pnictides, except that there are certain amounts of
Fe-vacancies in the iron square sublattice. The iron vacancies are
expected to order in some periodic superstructures, rather than to
distribute randomly within the FeSe layer, resulting in the normal
state insulating behavior \cite{fan11}. Indeed, the $\sqrt{5} \times
\sqrt{5}$ vacancy ordering pattern as shown in Fig.~\ref{f1},
corresponding to $x=0.4$, seems to be most stable as confirmed
in the neutron diffraction \cite{bao11,ye11,pom11} and
transmission electron microscopy \cite{wan11b} experiments.
In addition, a novel magnetic ordering pattern, i.e.,
block spin checkerboard (BSC) state has also been observed, with
an unexpected high transition temperature $T_N \sim 550$ K and a
large effective magnetic moment $\sim 3.31 \mu_B$/Fe.

Early theoretical explorations to the magnetic and electronic
structures of A$_{y}$Fe$_{1.6}$Se$_{2}$ based on the
first-principles calculations have revealed the BSC state as the
groundstate with effective iron moment $\sim 2.8\mu_B$/Fe$-3.4
\mu_B$/Fe, and a band gap $\sim 500$ meV at $y=0.8$ \cite{cao11,yan11}. Another
silent feature is a significant lattice contraction of the
fundamental iron blocks but without breaking the symmetry of lattice
structure. The observation leads to a microscopic consideration for
the magnetic structure based on the extended $J_1$-$J_2$ spin model
\cite{cao11}, where the BSC state could be the groundstate for
certain range of model parameters \cite{yu11}. This model involves
the nearest-neighbor (NN) and the next-nearest-neighbor (NNN)
exchange interactions, and captures the vacancy superstructure and
the lattice distortion in a minimal manner. Other models like the
$J_1$-$J_2$-$J_3$ spin model, which emphases the relaxation of
magnetic frustration by the third-nearest-neighbor exchange
interaction without \cite{fan12} or with \cite{hu12} a biquadratic
interaction term, can also account for the BSC state.

Experimentally, the vacancy order and the BSC state coexist with the
superconductivity in the A$_{y}$Fe$_{2-x}$Se$_{2}$ compounds for $y
\gtrsim 0.8$, $x\lesssim  0.4$. This raises heated debate on whether
the co-existence is an intrinsic property of a single phase
electronic structure \cite{liu11} or due to a phase separation
\cite{che11b}. Theoretically, the importance of the vacancy order
and the BSC state in the formation of the insulating phase as well
as superconductivity has been investigated
\cite{cao11c,yan11b,yu11b,zho11,wan11c,zha11,hu12,che11,jia12}.
However, the question that why the magnetic transition temperature
$T_{N}$ is much higher than the ones in other iron pnictides or
chalcogenides, has not been addressed so far by detailed
calculations.

This fundamental question is actually non-trivial. Firstly, it is
believed that unconventional superconductivity, which frequently
emerges near the border of antiferromagnetic phases such as in the
cuprates, heavy fermions, and iron pnictides/chalcogenides, is
usually related to the magnetic fluctuations with a characteristic
energy scale roughly proportional to the magnetic transition
temperature $T_{N}$ \cite{sca12}. But the present class of
intercalated iron chalcogenides (A$_{y}$Fe$_{2-x}$Se$_{2}$)
is an obvious exception. Secondly,
while the intralayer magnetic interactions $J_1$ and $J_2$ in
A$_{y}$Fe$_{2-x}$Se$_{2}$ are comparable to those in
iron pnictides \cite{wan11d} as well as in $A$Fe$_{2}$Se$_{2}$
without vacancies \cite{yan11c,cao11b}, the interlayer
magnetic coupling is very small \cite{cao11} and is not sufficient to
account for the enhancement of $T_{N}$. Thirdly, it is well-known
that uniformly and randomly distributed vacancies in
spin systems always lead to a reduction of transition temperatures
such as in the randomly site-diluted Ising model \cite{ken08}.
Therefore, detailed calculations are required to understand why and
how the opposite tendency could appear in the
A$_{y}$Fe$_{2-x}$Se$_{2}$ compounds.

In this paper, we consider this question by performing systematic
simulations on the finite-temperature magnetic phase transitions in
the extended $J_1$-$J_2$ spin model defined in a square lattice with
a $\sqrt{5} \times \sqrt{5}$ vacancy superstructure. This model has
a rich phase diagram consisting of numerous groundstates when
we tune the model parameters \cite{cao11,yu11}. We
then focus on three representative sets of model
parameters. Our goals are of two-folds: to clarify the nature of the
finite-temperature magnetic order-disorder phase transitions and to
understand the mechanism increasing the transition temperatures.
Though the relation between the magnetic transition
temperature $T_N$ and the onset of superconductivity is not
addressed, our study indicates that the enhancement of the critical temperature
may be irrelevant to the superconductivity.
In other words, our results favor the phase separation scenario \cite{che11b}.

According to the $\sqrt{5} \times \sqrt{5}$ vacancy
superstructure, the concentration of vacancies $p = 20\%$ is
considered, corresponding to $x=0.4$ in realistic
A$_{y}$Fe$_{2-x}$Se$_{2}$ compounds with the superconductivity and
high magnetic transition temperature. The specified spatial
distribution of vacancies as observed in experiments is the simplest
vacancy superstructure in a square lattice with the highest
symmetry, since all iron atoms are $3$-coordinated equivalently.

There are several technique difficulties to our numerical
investigations. For the systems with ($J_1,J_2$) couplings and
vacancies, sufficiently large system sizes are required in order
to find the true groundstate among numerous possible magnetic
configurations. For this purpose, the parallel tempering
technique based on Monte Carlo method, which was applied in studying
spin glass systems \cite{rom09}, is an appropriate approach. Once
the groundstate is determined, the magnetic order-disorder phase
transition at finite temperatures can be investigated by large-scale
Monte Carlo simulations. However, even for the frustrated Ising model
without vacancies, it is difficult to precisely determine
the order and the universality class of the phase transitions.\cite{yin09, jin12}
In this respect, the short-time dynamic approach \cite{zhe98,luo98} can be utilized.
Recent activities include various applications and developments
\cite{oze07, alb11} such as theoretical and numerical studies of the
Josephson-junction arrays \cite{gra05} and ageing phenomena
\cite{cal05,lei07}. Very recently, the depining transition and the
relaxation-to-creep transition in the domain-wall motion have been
investigated \cite{zho09,zho10,zho10a,qin12}. Usually,
two relaxation processes with ordered and disordered initial states
are considered in the short-time dynamic approach. However, the
groundstate of the present model is not simply ordered and
homogeneous, and the standard magnetization does not characterize
these relaxation dynamics. Methodologically and technically one
needs to develop new concepts.

With the parallel tempering technique and the short-time dynamic
approach, we are able to find the true groundstate as
well as to accurately determine the transition temperature and
critical exponents. In Sec. II, the model and scaling analysis are
described, and in Sec. III, the numerical results are presented.
Finally, Sec. IV is devoted to the conclusions.

\section{Model and scaling analysis}

\subsection{Model and methods}
\begin{figure}[ht]
\epsfysize=6.5cm \epsfclipoff \fboxsep=0pt
\setlength{\unitlength}{1.cm}
\begin{picture}(10,6)(0,0)
\put(1.5,-0.3){{\epsffile{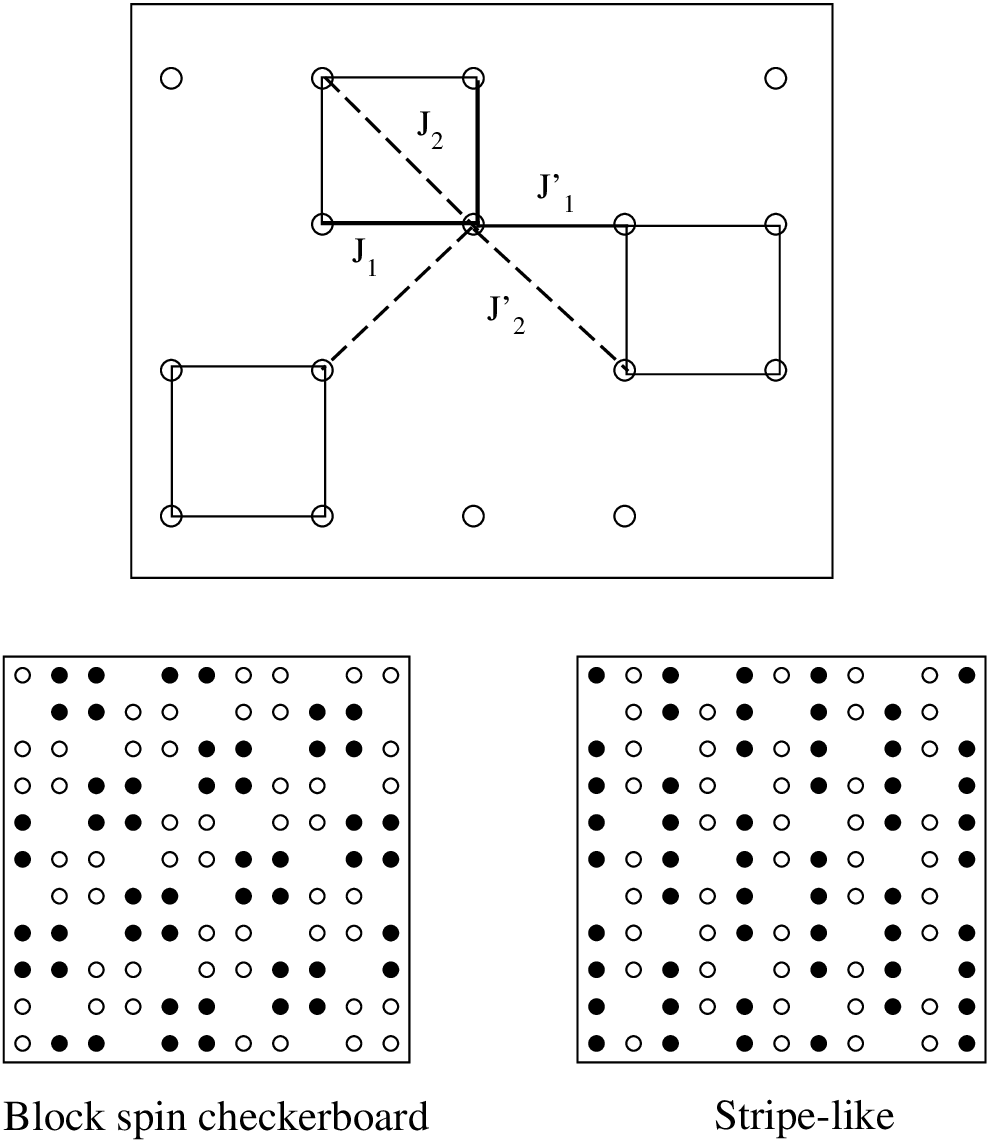}}}
\end{picture}

\caption{The magnetic structure of A$_{y}$Fe$_{2-x}$Se$_{2}$ from
top-view is displayed. In the upper panel, the solid squares
connecting the circles indicate the fundamental blocks with four Fe
atoms at the corners. The proposed magnetic couplings $(J_1, J'_1)$
with solid lines and $(J_2, J'_2)$ with dashed lines represent the
NN and NNN couplings, respectively. In the lower panel, two magnetic
configurations, the block spin checkerboard and stripe-like
antiferromagnetic states, are shown by open ($S_i = 1$) and solid
circles ($S_i = -1$).} \label{f1}
\end{figure}

The extended $J_1$-$J_2$ model is the extension of standard
$J_1$-$J_2$ model defined on the Fe-square lattice of iron pnictides
\cite{si08} to the situation with a Fe-vacancy superstructure. Here,
the distance between two NN vacancies is $\sqrt{5}$ in unit of the
distance between two NN iron sites. Owing to the
$\sqrt{5}\times\sqrt{5}$ vacancy superstructure, the whole lattice
consists of the fundamental minimal square blocks containing four
iron sites. Due to the symmetry invariant lattice distortion, the
intrablock and interblock interactions could be different
\cite{cao11}. The model is then defined by the following
Hamiltonian,
\begin{eqnarray}
H &= &\sum_{n,\alpha}(J_1S_{n,\alpha}S_{n,\alpha+1} +
J_2S_{n,\alpha}S_{n,\alpha+2}) \nonumber \\
&+&\sum_{n,\alpha}J_1'S_{n,\alpha_{\delta}}S_{n+\delta,\alpha_{\delta}}
+\sum_{n,\alpha}J_2'(S_{n,\alpha_{\delta}}S_{n+\delta,\alpha_{\delta}+1}\nonumber\\
&+&S_{n,\alpha_{\delta}}S_{n+\delta-1,\alpha_{\delta}-1}),
\label{equ10}
\end{eqnarray}
where $n$ denotes the block index, $n + \delta$ is short for the
nearest-neighboring block, $\alpha$ is the site index which goes
from $1$ to $4$, and $\alpha_{\delta}$ selects the site connecting
to the nearest-neighboring block. $J_1$ and $J_1'$ ($J_2$ and
$J_2'$) are the NN (NNN) couplings of intra- and interblock, as
shown in Fig.~\ref{f1}. Since an almost saturated magnetic moment at
the iron site is reported \cite{bao11}, the quantum effect is
suppressed at finite temperatures due to the large local spin. So
the corrections due to quantum fluctuations to the magnetic
properties at finite temperatures can be safely neglected.
Therefore, the spin is treated as a classical Ising spin, i.e.,
$S_{n,\alpha} = \pm 1$. The specific value of $|S|$ is not crucial
in the following discussions.

The general phase diagram of this classical $J_1$-$J_2$
spin model is very complicated. A previous Monte Carlo study
suggests a phase diagram with several specified groundstates for the
vector version of this model with certain model parameters
\cite{yu11}. For our purpose, we consider more simplified cases by
fixing the non-zero coupling strengths $|J|=1$. Explicitly, three
representative sets of model parameters are considered,
\begin{equation}
\left \{
   \begin{array}{lllll}
\mbox{Case-I:}   & \quad  & J_1 = J_2=-1, &~& J_1'=J_2'=1 \\
\mbox{Case-II:}  & \quad  & J_1 = J_2=1 , &~& J_1'=J_2'=1 \\
\mbox{Case-III:} & \quad  & J_1 =-1,J_2=0, &~& J_1'= 1,J_2'=0 \\
   \end{array}
   \right..
\label{equ20}
\end{equation}
Here, Case-I involves the ferromagnetic intrablock NN and NNN
couplings and antiferromagnetic interblock NN and NNN couplings,
while all NN and NNN couplings in Case-II are antiferromagnetic.
Case-III is a simplification of Case-I, i.e., without the NNN
interactions. First-principles calculations suggest that in
realistic A$_{y}$Fe$_{2-x}$Se$_{2}$ compounds, $J_1$, $J_2$, and
$J_1'$ are ferromagnetic, while $J'_2$ is antiferromagnetic
($J_1=-43$ meV, $J_2 = -4.5$ meV, $J_1'=-14.5$ meV,
$J_2' = 19$ meV). However, $J'_2$ dominates over $J'_1$ not only
because $|J'_2|$ is larger than $|J_1'|$, but also because the
number of interblock NNN sites is twice of the interblock NN sites
\cite{cao11}. Thus the sign of $J_1'$ is not crucial in this case.
The origin of the ferromagnetic couplings is a combined
effect of Hund's rule coupling and short-ranged hopping integrals
(of Fe $3d$-orbitals and Se $4p$-orbitals) which are enhanced by the
vacancy-induced lattice contraction. Notice that when no iron
vacancies appear as such in $A$Fe$_2$Se$_{2}$ \cite{cao11b,yan11c},
both NN and NNN interactions are antiferromagnetic as in other iron
pnictides.

In the following, we shall find the magnetic
configurations of the groundstate and investigate the
finite-temperature magnetic order-disorder transitions in each of
the three cases with Monte Carlo simulations. To overcome the
critical slowing down around the phase transition, we adopt the
short-time dynamic approach. Two relaxation processes,
i.e., those starting from the groundstate (ordered state) and
high-temperature state (disordered state) are considered. To
extract the transition temperature and critical exponents, it is
more efficient to study the dynamic relaxation starting from the
groundstate. Fig.~\ref{f1} shows two typical magnetic
configurations, corresponding to the magnetic ordering patterns of
the BSC state and the stripe (or collinear) antiferromagnetic state
(denoted by SAFM), as observed in realistic
A$_{y}$Fe$_{2-x}$Se$_{2}$ compounds and iron pnictides
\cite{bao11,cru08}. We will show that the groundstates of Case-I and
Case-II are the BSC and SAFM states, respectively.

For a simple model, such as the Ising model with only
NN couplings on a square lattice, it is straightforward to obtain
the groundstate magnetic configuration according to the symmetry.
However, the vacancy order and frustrated antiferromagnetic NNN
interactions in the extended $J_1$-$J_2$ model make this task
difficult. Large-scale simulations implemented by the parallel tempering
algorithm are then performed to find the true groundstate. The
details of the parallel tempering algorithm can be found in
Ref.\cite{rom09}, and the main idea is briefly illustrated below. In
this algorithm, $m$ parallel replicas are analyzed, each of which is
performed independently at a fixed temperature $T_j$ ($T_1 \leq T_j
\leq T_m$). Following the reference, we fix $T_1 = 0.1, T_m = 1.6$
and set $T_{j+1} - T_j=(T_m-T_1)/(m-1)$. In order to avoid the
situation where replicas at low temperatures get stuck in local
minima, one can swap the configurations of two randomly selected
temperatures $T_j$ and $T_{j'}$. Starting from a random initial
condition, a standard Monte Carlo dynamics is performed in each
replica, and a trial exchange of two configurations $X_j$ and
$X_{j'}$ (corresponding to the $j$th and $j'$th replicas) is
attempted periodically, and accepted with the probability
\begin{equation}
W(X_{j}, K_j|X_{j'}, K_{j'}) = \left \{
   \begin{array}{lll}
   \exp(-\Delta),  & \quad &  \mbox{for $\Delta > 0$} \\
   1,    & \quad & \mbox{for $\Delta \leq 0$}
   \end{array}
   \right..
\label{equ30}
\end{equation}
Where $\Delta = - (K_j-K_j')(H_j-H_j')$ is defined with the inverse
temperature $K_j=1/T_j$ and Hamiltonian energy $H_j$. For
convention, we restrict the replica exchange to the case $j'=j+1$.
As time evolves, the magnetic configuration at the lowest
temperature approaches to the groundstate.

After preparing the groundstate as the initial state, we update the
spins with the heat-bath algorithm.  Our simulations are performed
with lattice sizes $L=250, 500,$ and $1000$, up to
$t_{max}=25~600$ Monte Carlo step (MCS). Here MCS is defined by $L
\times L$ single-spin-flips attempts. Different
updating schemes, such as the sequential sweep and random sweep, are
considered, and yield the same results. Periodic boundary
conditions are used along the $x$ and $y$ directions, respectively.
For each case, more than $16~000$ samples are
performed for average. Errors are estimated by dividing the samples
into three or four subgroups. If the fluctuation of the curve in the
time direction is comparable with or larger than the statistical
error, it will be taken into account.

To investigate the dynamic relaxation, the pseudo-magnetization
$M(t)\equiv M^{(1)}(t)$ and its second moment $M^{(2)}(t)$ are
introduced by the projection to the groundstate,
\begin{equation}
M^{(k)}(t) = \frac{1}{L^{2k}} \left \langle \left [\sum_{
i}S_{i}(t)X_{i}\right ]^k \right \rangle,  \quad k = 1,2,
\label{equ40}
\end{equation}
where $S_i(t)$ is the spin at the time $t$ on the lattice site $i$,
$X_i$ is the one from the groundstate, $L$ is the lattice size, and
$\langle\cdots\rangle$ represents the thermal average,
estimated by the average over samples with different random numbers
and initial conditions. The pseudo-magnetization $M(t)$ plays a
role as the order parameter of the magnetic transition. When the
groundstate is degenerate, a computationally convenient
root-mean-square order parameter is introduced \cite{yin09}. Other
important observables are the susceptibility $\chi(t)$ and Binder
cumulant $U(t)$,
\begin{eqnarray}
\chi(t)& \sim & M^{(2)}(t) - M(t)^{2}, \nonumber \\
U(t)& \sim &\chi(t)/M(t)^{2}. \label{equ60}
\end{eqnarray}
For the dynamic relaxation starting from the disordered state, the
spatial correlation function $C(r,t)$ and two-time correlation
function $A(t,t')$ are measured,
\begin{eqnarray}
C(r,t)& = & \frac{1}{L^d}\sum_{i} \left \langle S_i(t)S_{i+r}(t)\right \rangle, \nonumber \\
A(t,t')& = & \frac{1}{L^d} \sum_{i} \left \langle
S_i(t')S_{i}(t)\right \rangle, \label{equ70}
\end{eqnarray}
where $r$ is the spatial distance, $t'$ is the waiting time, and
$d=2$ is the spatial dimension.

\subsection{Scaling analysis}
The magnetic order-disorder transition at finite temperatures in the
present model is of the second order, compatible with the magnetic
transitions in the A$_{y}$Fe$_{2-x}$Se$_{2}$ compounds \cite{bao11},
where no lattice structural transition is accompanied with the
magnetic ordering except for the iron vacancy ordering stability
taking place at an elevated temperature $T_V\sim 580$ K. Hence, one
expects that the order parameter $M(t)$ should obey the dynamic
scaling form, after a microscopic time scale $t_{mic}$ \cite{zhe98},
\begin{equation}
M^{(k)}(t, \tau, L) = t^{-k\beta/\nu z} \widetilde{M}(t^{1/\nu
z}\tau, t^{1/z}/L), \label{equ80}
\end{equation}
here $\beta$ and $\nu$ are the static exponents, $z$ is the dynamic
exponent, and $\tau = (T - T_c) / T_c$ is the reduced temperature.
$T_c$ denotes the transition temperature which can be either Curie
temperature $T_C$ in the ferromagnetic transition or N\'eel
temperature $T_N$ in the antiferromagnetic transition. On the right
side of the equation, the overall factors $t^{-k\beta/\nu z}$
indicates the scaling dimension of $M(t)$, and the scaling function
$\widetilde{M}(t^{1/\nu z}\tau, t^{1/z}/L)$ represents the scale
invariance of the dynamic system. For a sufficiently large lattice
and in the short-time regime, the nonequilibrium spatial correlation
length $\xi(t)\sim t^{1/z}$ is much smaller than the lattice size
$L$. Therefore, the finite-size effect is negligible, and a power
law behavior is expected at $\tau = 0$,
\begin{equation}
M(t) \sim t^{-\beta /\nu z}. \label{equ90}
\end{equation}
With Eq.~(\ref{equ80}), the precise location of the transition
temperature $T_c$ is determined by searching for the best power-law
behavior of $M(t,\tau)$, and the critical exponent $1/\nu z$ is
measured from the time derivative of $\ln M(t, \tau)$.

For the susceptibility $\chi(t)$ and Binder cumulant $U(t)$, the
scaling behaviors are different. Since $\xi(t)$ is small, the
spatially correlating terms $\langle S_1 X_1 S_2 X_2 \rangle$ with
$|r_1-r_2|> \xi(t)$ can be neglected. In other words, one of the two
summations over $r_1$ and $r_2$ in $M^{(2)}(t)-M(t)^{2}$ is
suppressed. It then leads to the finite-size behaviors $\chi(t),
U(t) \sim L^{-d}$. Together with Eqs.~(\ref{equ80}) and
(\ref{equ90}), one may derive the scaling forms,
\begin{eqnarray}
\chi(t) & \sim & t^{\gamma /\nu z} / L^d, \nonumber \\
U(t)& \sim & t^{d/z} / L^d, \label{equ100}
\end{eqnarray}
with the scaling law $\gamma / \nu = d - 2 \beta / \nu$.

For the dynamic relaxation starting from the disordered state, the
correlation functions $C(r,t)$ and $A(t,t')$ should obey
\begin{eqnarray}
C(r,t)& \sim & t^{-2\beta/\nu z} \widetilde{C}\left( r / \xi(t) \right), \nonumber \\
A(t,t')& \sim & t'^{-2\beta/\nu z}\widetilde{A}\left( \xi(t) /
\xi(t')\right), \label{equ110}
\end{eqnarray}
where $\widetilde{C}(s)$ and $\widetilde{A}(q)$ are the scaling
functions with $s=r/\xi(t)$ and $q=\xi(t)/\xi(t')$. Together with
Eqs.~(\ref{equ70}) and (\ref{equ110}), one may derive the scaling
form of the integral $S(t) = \int C(r,t) dr$,
\begin{equation}
S(t) \sim t^{ (d_0 - 2\beta / \nu)/ z}, \label{equ120}
\end{equation}
where $d_0$ denotes the dimension of the integration. For a
sufficiently large lattice and at the critical point $T_c$, a
surprising increasing behavior of the pseudo-magnetization $M(t)$ is
observed,
\begin{equation}
M(t) \sim m_0t^{\theta}. \label{equ130}
\end{equation}
Here $m_0$ is the initial magnetization, and $\theta$ is a local
critical exponent, reflecting the effect of the initial condition
\cite{jan89}.

\section{Monte Carlo simulations}
As shown in Fig.~\ref{f2}(a), the second moment of
pseudo-magnetization is displayed in the parallel tempering process
at the lowest-temperature replica for Case-I and Case-II. The
lattice size $L=200$ and replica number $m=10$ are used. As time
grows, the curves approach to the unit. It indicates that the BSC
and SAFM states, displayed in Fig.~\ref{f1}, are the true
groundstates for Case-I and Case-II, respectively. As a test, we
also consider a set of model parameters by re-scaling the couplings
obtained from the first-principles calculations \cite{cao11}. In
this case, more than $1~000$ samples are performed, up to
$t_{max}=1~000~000$ MCS. All of them evolve to the BSC
state, reflecting the fact that there is an extended region of the
BSC state in the groundstate phase diagram of the model \cite{yu11}.

\subsection{Dynamic relaxation from groundstate}

With Monte Carlo simulations, the dynamic relaxation starting from
the groundstate is investigated. In Fig.~\ref{f2}(b), the time
evolution of the pseudo-magnetization $M(t)$ in Case-I is displayed
for different inverse temperatures $K=1/T$ with the lattice size
$L=500$. The curve drops rapidly down for smaller $K$, while
approaches a constant for larger $K$. Searching for the best
power-law behavior, the critical point $K_c=0.27595(3)$ is
determined accurately. According to Eq.~(\ref{equ90}), one measures
the exponent $\beta/\nu z = 0.0585(6)$ from the slope of the curve
at $K_c$. Additional simulations with $L=250$ and $L=1000$ confirm
that the finite-size effect is already negligibly small. For
comparison, the dynamic behavior in Case-II is also studied, and the
critical point $K_c=0.8148(1)$, the exponent $\beta / \nu z =
0.0556(3)$ are derived.

\begin{figure}[ht]
\epsfysize=6.5cm \epsfclipoff \fboxsep=0pt
\setlength{\unitlength}{1.cm}
\begin{picture}(10,6)(0,0)
\put(-3.0,-0.3){{\epsffile{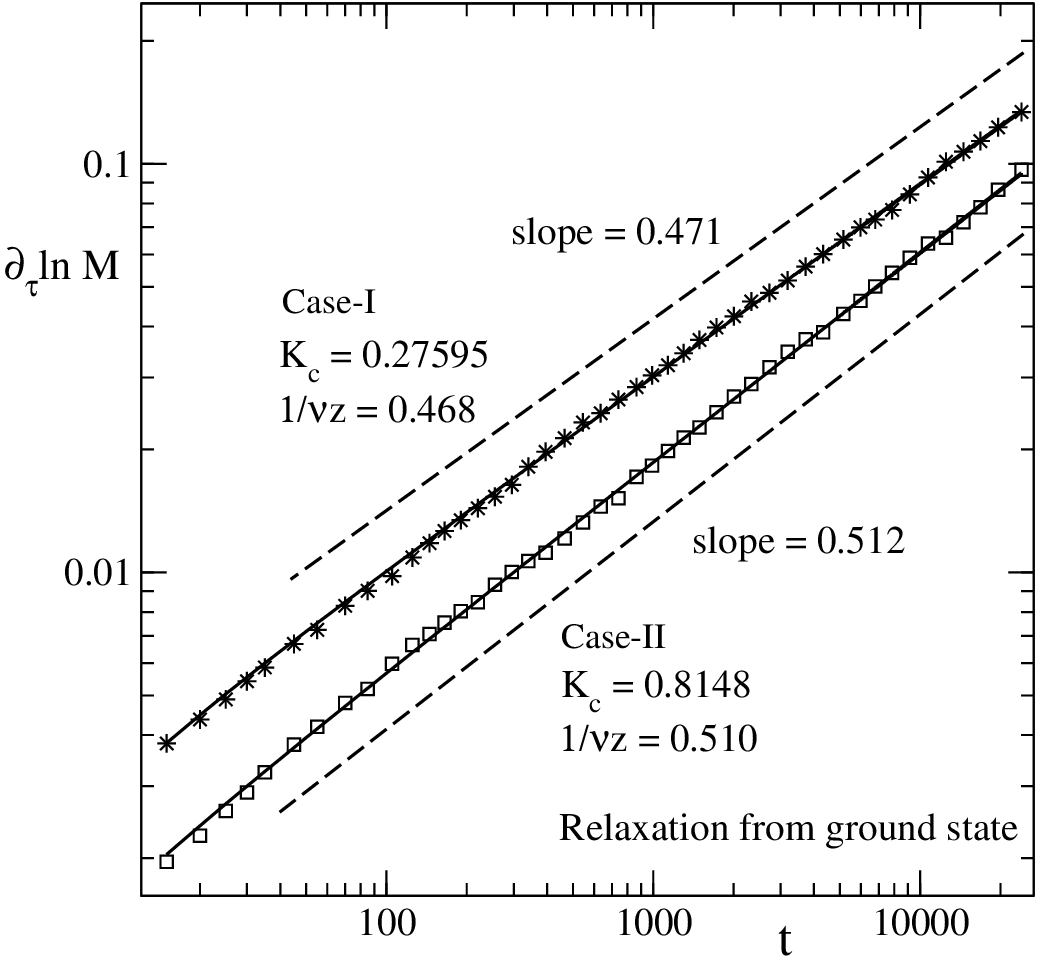}}}\epsfysize=6.5cm
\put(5.2,-0.3){{\epsffile{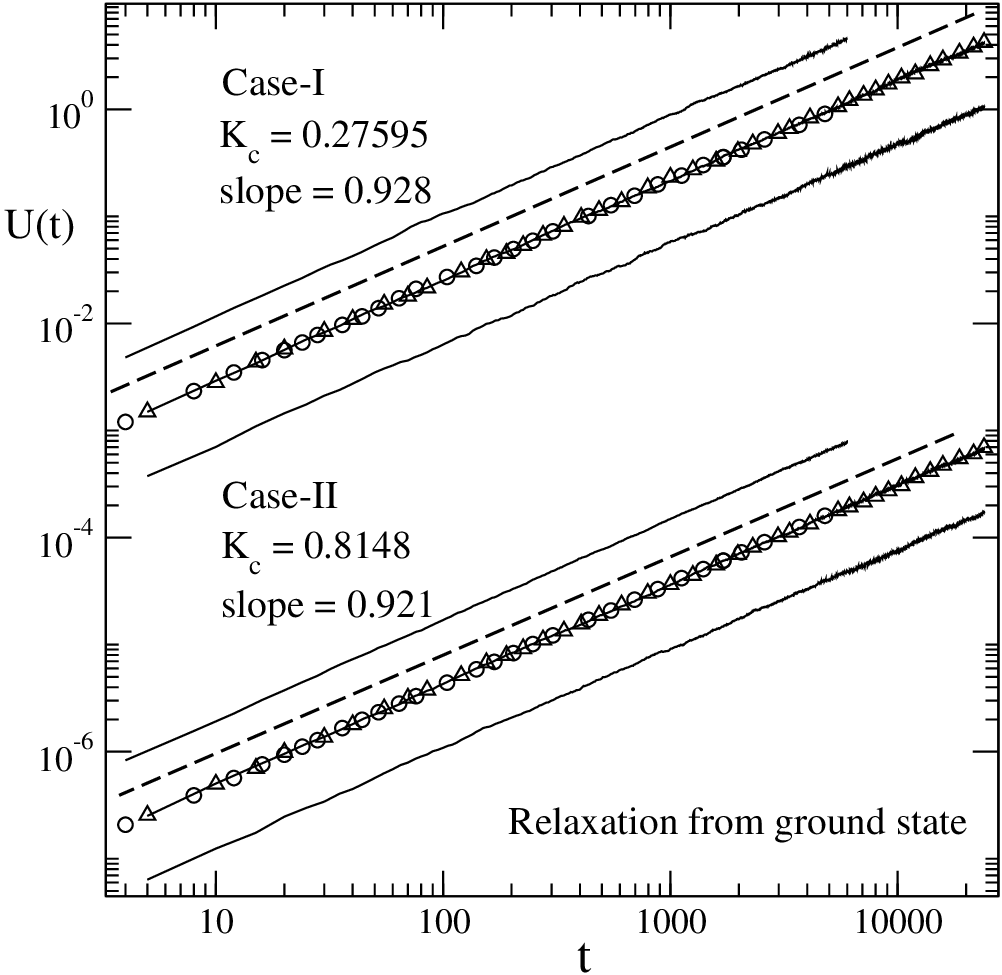}}}
\end{picture}

\hspace{1.0cm}\footnotesize{(a)}\hspace{8.0cm}\footnotesize{(b)}
\caption{(a) The logarithmic derivative of the pseudo-magnetization
$M(t, \tau)$ is displayed at $K_c$ for Case-I (stars) and
Case-II (open squares). Dashed lines represent
power-law fits, and solid lines indicate the fits with power-law
correction. \quad(b) The Binder cumulant $U(t)$ is plotted with
solid lines on a double-log scale for different lattice size $L=250,
500$ and $1000$ (from above). According to Eq.~(\ref{equ100}), data
collapse is demonstrated at a fixed lattice size $L=500$. Open
circles and triangles correspond to $L=250$ and $1000$,
respectively.}\label{f3}
\end{figure}

In order to approximate the differentiation of $\ln M(t, \tau)$, the
simulations at temperatures in the vicinity of the critical point
are performed. In Fig.~\ref{f3}(a), a power-law behavior of the
curves is observed but with certain corrections to scaling at the
early times. A direct measurement from the slope gives the exponents
$0.471(5)$ and $0.512(3)$ for Case-I and II, respectively. After
introducing a power-law correction to scaling, $\partial_{\tau} \ln
M(t) \sim t^{1/\nu z} (1 + c / t)$  \cite{lei07}, one can fit the
numerical data extending to rather early times. It yields $1/\nu z =
0.468$ in Case-I, and $0.510$ in Case-II.

In Fig.~\ref{f3}(b), the time evolution of the Binder cumulant
$U(t)$ is plotted at $K_c$ for these two cases. The possible
finite-size behavior is also investigated with different lattice
sizes $L=250, 500$ and $1000$, and data collapse is observed
according to Eq.~(\ref{equ100}). From the slope, one measures the
exponent $d /z = 0.928(5)$ in Case-I, and $0.921(5)$ in Case-II.

Finally, according to the measurements of $\beta/\nu z$, $1/\nu z$,
and $d/z$, we calculate the individual exponents
$\beta=0.125(2),\nu=1.00(1), z=2.16(1)$ in Case-I, and
$\beta=0.109(1),\nu=0.90(1), z=2.17(1)$ in Case-II.

\subsection{Dynamic relaxation from disordered state}

\begin{figure}[ht]
\epsfysize=6.5cm \epsfclipoff \fboxsep=0pt
\setlength{\unitlength}{1.cm}
\begin{picture}(10,6)(0,0)
\put(-3.0,-0.3){{\epsffile{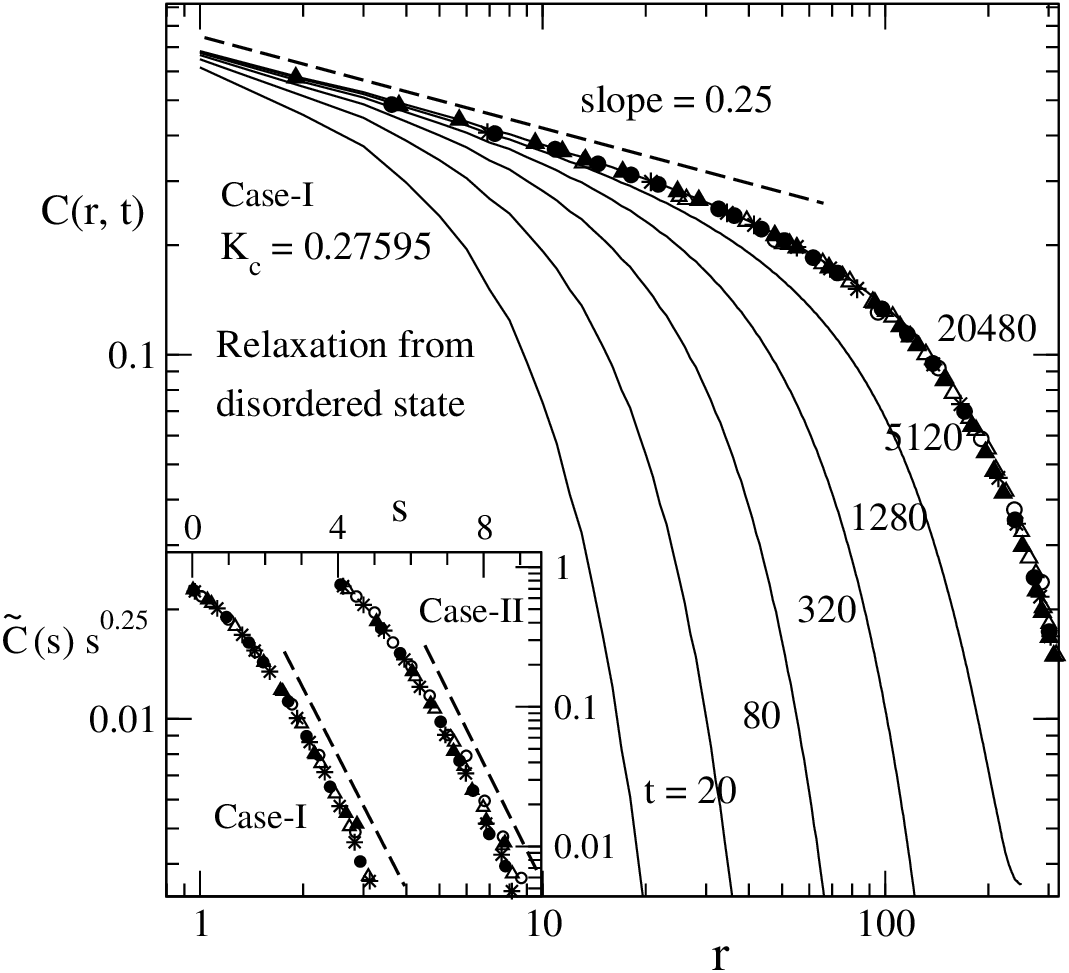}}}\epsfysize=6.5cm
\put(5.2,-0.3){{\epsffile{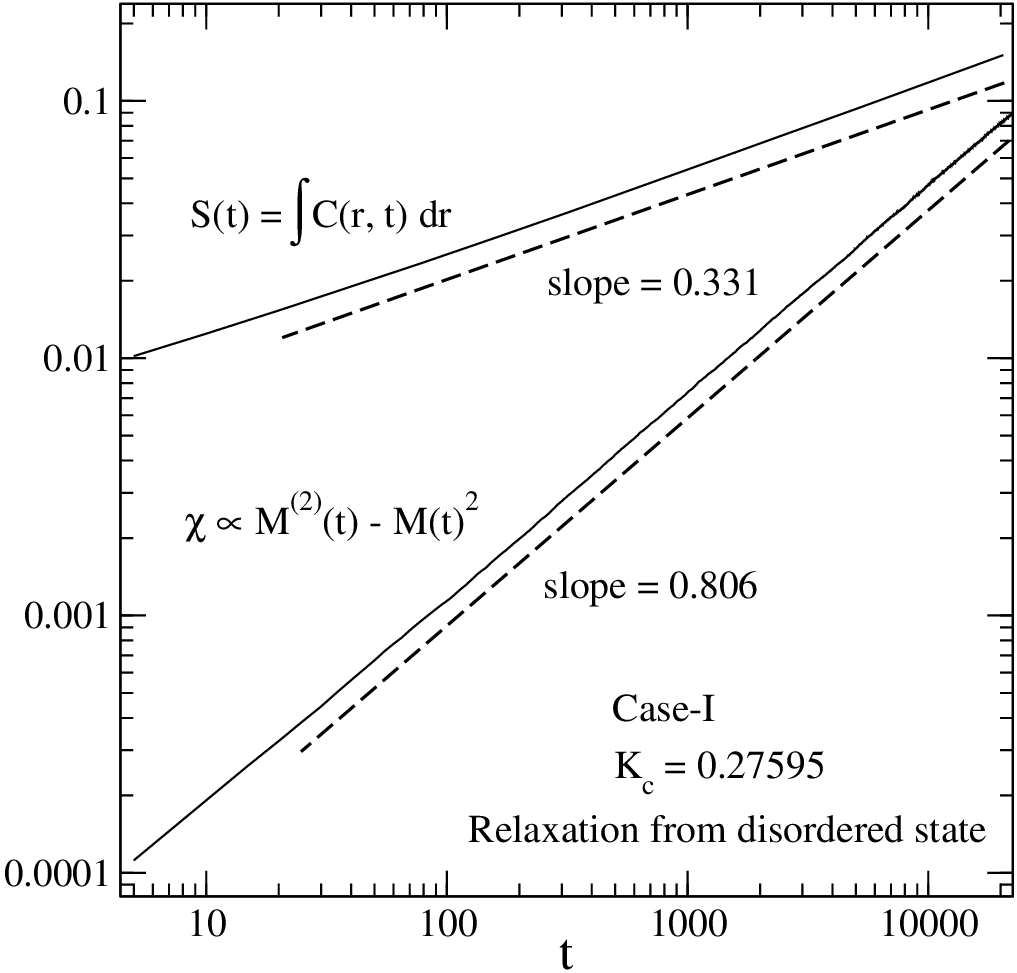}}}
\end{picture}

\hspace{1.0cm}\footnotesize{(a)}\hspace{8.0cm}\footnotesize{(b)}
\caption{(a) The spatial correlation function $C(r, t)$ is displayed
on a log-log scale. Data collapse is demonstrated at a fixed
$t=20480$ MCS. Open circles, open triangles, stars, solid circles,
and solid squares correspond to $t=20, 80, 320, 1280$, and $5120$,
respectively. In the inset, the scaling function
$\tilde{C}(s)s^{0.25}$ is shown on a linear-log scale.  For clarity,
the curve of Case-II is shifted right. \quad(b) Dynamic relaxation
of $\chi(t)$ and $S(t)$ are plotted with solid lines for Case-I.
Dashed lines represent power-law fits.}\label{f4}
\end{figure}

Now we turn to the dynamic relaxation starting from the disordered
state at the critical temperature $T_c$. In Fig.~\ref{f4}(a), the
spatial correlation function $C(r,t)$ is displayed for Case-I as a
function of distance $r$ at different time $t$. To confirm the
scaling behavior of $C(r,t)$, for example, we fix $t'=20480$ MCS,
and rescale $r$ to $(t'/t)^{1/z}r$ and $C(r,t)$ to
$(t'/t)^{-2\beta/\nu z}C(r,t)$. Data of different $t$ nicely
collapse to the curve of $t'$ with the exponents $\beta / \nu z
=0.0585$ and $z=2.16$ as input. A power-law decay is then observed
at small $s=r/\xi(t)$ with the slope $2\beta/\nu=0.25(1)$. In order
to extract the characteristic of the scaling function,
$\widetilde{C}(s)s^{0.25}$ against $s$ is plotted in the inset. For
large $s$ (e.g. $s\geq 2$), an exponential behavior is detected,
indicating the scaling form,
\begin{equation}
\widetilde{C}(s) \sim s^{-2\beta/\nu}\exp(-\alpha s). \label{equ140}
\end{equation}
Together with Eqs.~(\ref{equ110}) and (\ref{equ140}), one may derive
the critical behavior of the spatial correlation function $C(r,t)$
in the limit $r/\xi(t) \rightarrow \infty$,
\begin{equation}
C(r,t) \sim \frac{1}{r^{2\beta/\nu}}\exp \left(-\alpha r/ \xi(t)
\right), \label{equ150}
\end{equation}
here $\xi(t)\sim t^{1/z}$ is the spatial correlation length.

In Fig.~\ref{f4}(b), the integrated correlation function $S(t)$ is
displayed for Case-I, and the exponent $(d_0 - 2 \beta/\nu)/ z =
0.331(8)$ is estimated from the slope, according to
Eq.~(\ref{equ120}). The dimension $d_0=0.97(2)$ is calculated, very
close to $1$. Similarly, a power law behavior is also observed for
the susceptibility $\chi(t)$ with the slope $\gamma/\nu z=0.806(4)$.
It yields the exponent $\gamma =1.72(2)$. While similar measurement
for Case-II yields a different value of $\gamma =1.58(2)$.

\begin{figure}[ht]
\epsfysize=6.5cm \epsfclipoff \fboxsep=0pt
\setlength{\unitlength}{1.cm}
\begin{picture}(10,6)(0,0)
\put(-3.0,-0.3){{\epsffile{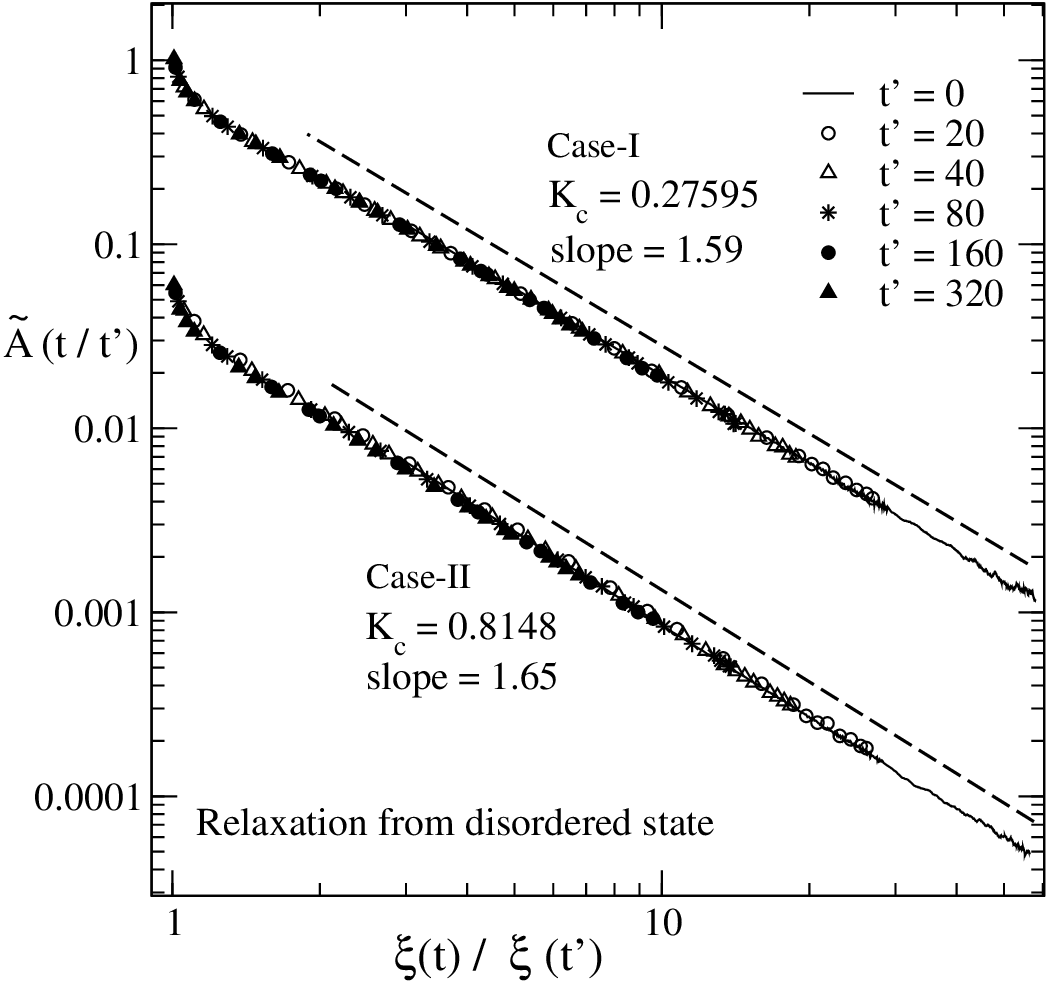}}}\epsfysize=6.5cm
\put(5.2,-0.3){{\epsffile{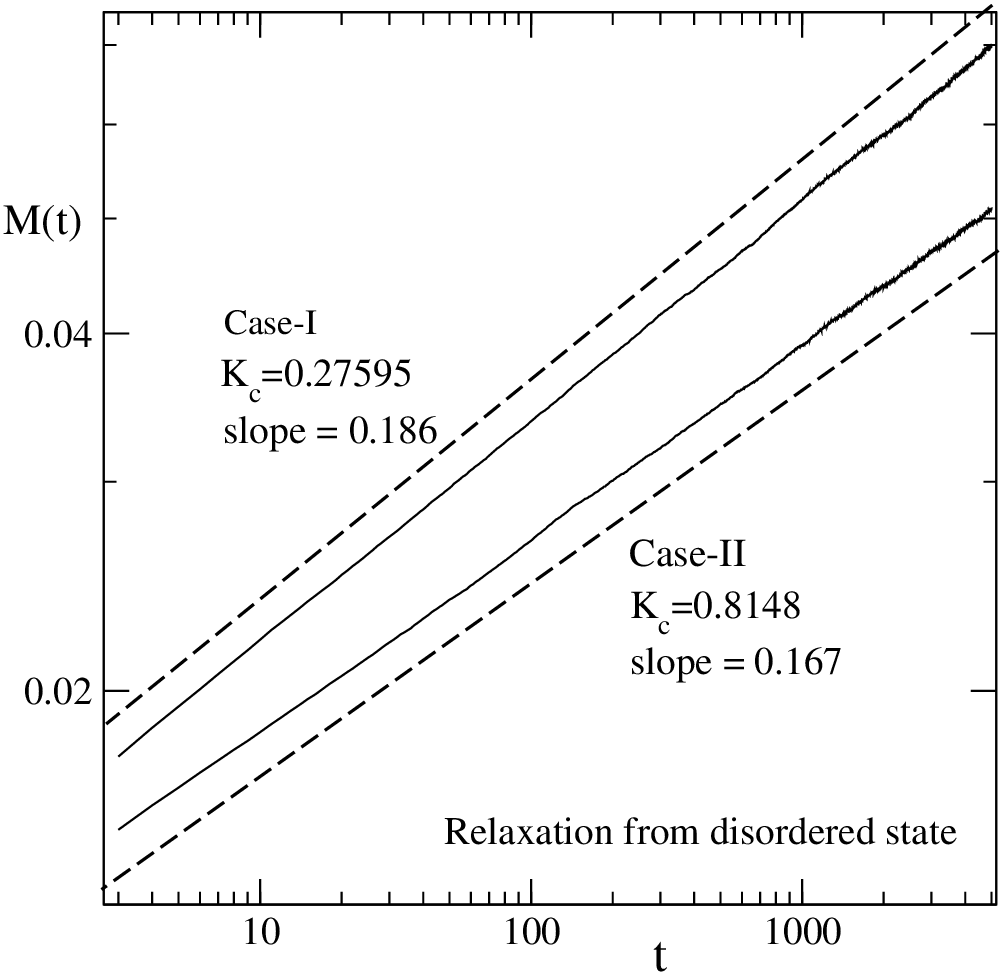}}}
\end{picture}

\hspace{1.0cm}\footnotesize{(a)}\hspace{8.0cm}\footnotesize{(b)}
\caption{(a) The scaling function $\tilde{A}(t/t')$ against
$\xi(t)/\xi(t')$ is displayed for Case-I and II on a double-log
scale. According to Eq.~(\ref{equ110}), data collapse is observed
for different waiting time $t'$. \quad(b) The time evolution of
$M(t)$ is plotted for Case-I and II with an initial magnetization
$m_0=0.01$. The lattice size is $L=1000$. In both (a) and (b),
dashed lines show power-law fits.}\label{f5}
\end{figure}

The scaling behavior of the two-time correlation function represents
a kind of ageing phenomena \cite{cal05,lei07}. According to
Eq.~(\ref{equ110}), the scaling function $\widetilde{A}(t/t')$ is
plotted in Fig.~\ref{f5}(a), as a function of $q=\xi(t)/\xi(t')$.
Obviously, data for different waiting time $t'$ collapse onto a
master curve which exhibits a power law decay in the
large $q$ regime (e.g. $q \geq 2$). It indicates that the scaling
function $\widetilde{A}(q)$ takes the form
\begin{equation}
\widetilde{A}(q) \sim q^{-\lambda}, \label{equ160}
\end{equation}
with the scaling law $\lambda = d - \theta z$. According to the
formula, the critical exponents $\lambda = 1.59(1)$ and $1.65(1)$
are estimated for Case-I and II, respectively.

Finally, a surprising increase of the pseudo-magnetization $M(t)$ is
displayed in Fig.~\ref{f5}(b) with the lattice size $L=1000$. From
the slope of the curve, one measures the critical exponent $\theta$.
Strictly speaking, $\theta$ is defined at the limit $m_0 \rightarrow
0$. However, { practical measurement} at this limit is
not possible. In this work, the initial magnetization $m_0=0.01$ is
prepared, which is believed to be small enough. It yields the
exponent $\theta=0.186(2)$ in Case-I, larger than the corresponding
value $0.167(1)$ in Case-II.

\subsection{Discussion}

All the measurements of the transition temperature and critical
exponents are summarized in Table~\ref{t1}, in comparison with those
of the $2D$ square Ising model without vacancies. In general, the
vacancies would lead to a reduction in the transition
temperature \cite{ken08}. Magnetic frustration induced by the
antiferromagnetic NNN interaction will also decrease the transition
temperature \cite{yin09}.  For example, the calculated critical
temperature $T_c = 1/K_c=1.2273(1)$ in Case-II is much lower than
the one $2.2692$ in the $2D$ Ising model. The reduction of $T_c$
should be due to both vacancies and magnetic frustration. However, a
dramatic enhancement in the transition temperature, $T_c=3.6238(4)$,
is found in Case-I. This value is much larger than that of the $2D$
Ising model, and almost three times as large as that in Case-II.
Differences between Case-I and Case-II are also observed in
individual critical exponents $\beta, \nu, \gamma, \theta$, and
$\lambda$ which differ by about $10\%$. It suggests that the
magnetic transitions in Case-I and Case-II are not in the same
universality class. Further comparison shows that the former belongs
to the Ising universality class, while the latter does not.
Interestingly, the ratios $\beta/\nu, \gamma/\nu$, and
the dynamic exponent $z$ in Case-II agree well with the
corresponding values of the $2D$ Ising model. It supports the
dynamic generalization of the "weak universality" hypothesis
proposed by Suzuki \cite{tan87, yin09}, where only the reduced
critical exponents $\beta/\nu, \gamma/\nu$ and $z=\Delta/\nu$ are
universal, irrelevant to the details of the interactions.

\begin{figure}[ht]
\epsfysize=6.3cm \epsfclipoff \fboxsep=0pt
\setlength{\unitlength}{1.cm}
\begin{picture}(10,6)(0,0)
\put(-3.0,-0.3){{\epsffile{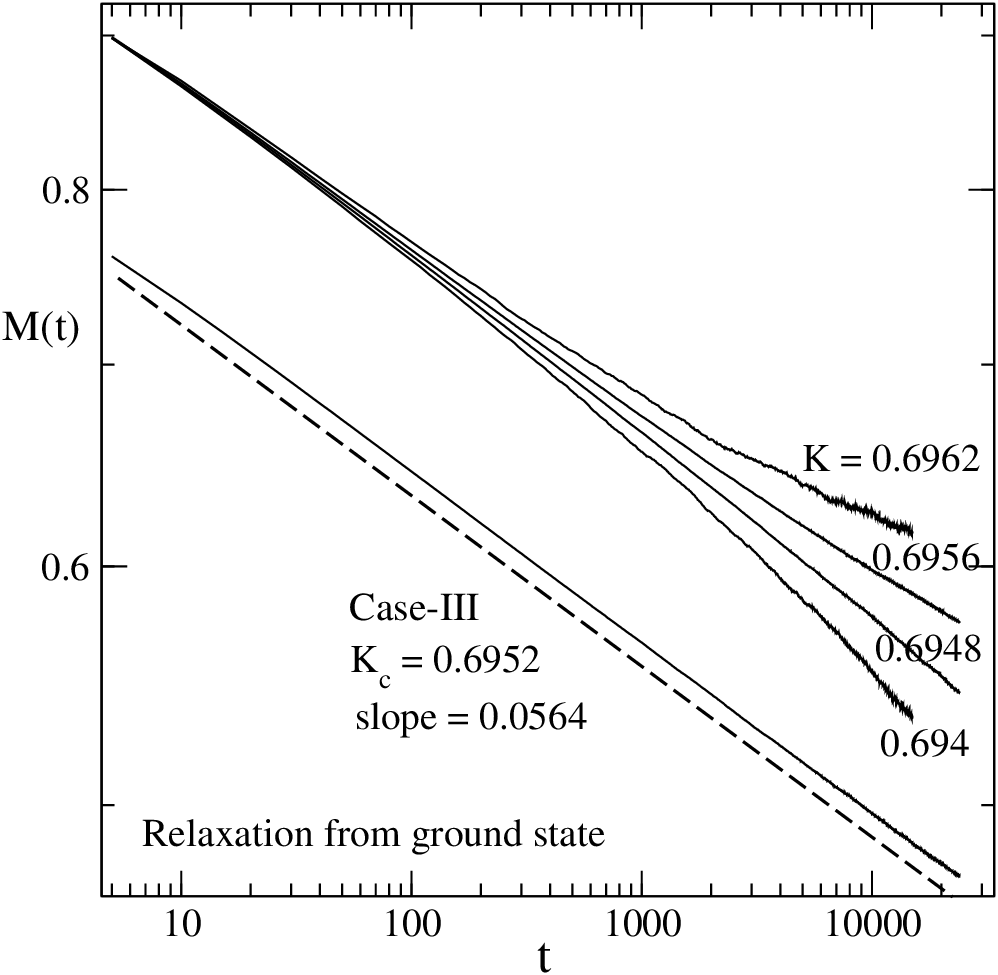}}}\epsfysize=4.0cm
\put(4.8,-0.3){{\epsffile{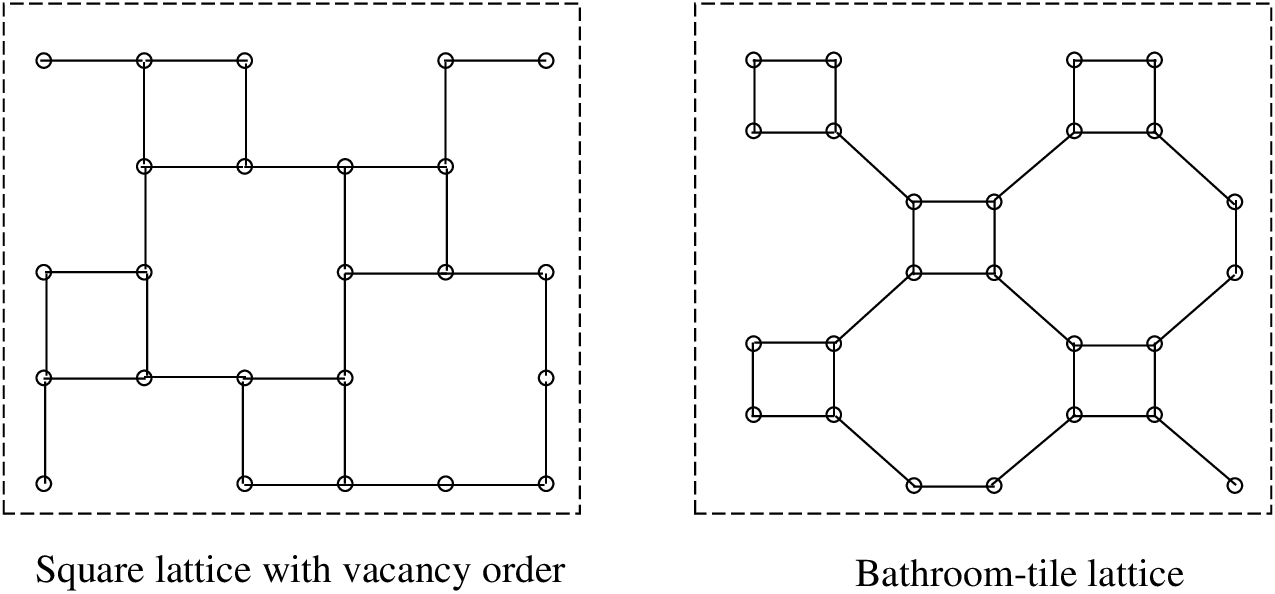}}}
\end{picture}

\hspace{1.0cm}\footnotesize{(a)}\hspace{8.0cm}\footnotesize{(b)}
\caption{(a) $M(t)$ in Case-III is plotted for different
temperatures. For clarity, the curve at $K_c$ is shifted down.
Dashed line shows a power-law fit.\quad(b) The square lattice with a
$\sqrt{5} \times \sqrt{5}$ vacancy superstructure and the
bathroom-tile lattice are shown within the dashed squares. A same
topological structure is revealed.}\label{f6}
\end{figure}

In order to understand above results, a simpler example, Case-III
defined in Eq.~(\ref{equ20}), is investigated. Using the parallel
tempering algorithm, the BSC state is confirmed as the groundstate,
too. In Fig.~\ref{f6}(a), the inverse transition temperature
$K_c=0.6952(1)$ and the exponent $\beta/\nu z = 0.0564(4)$ are
measured from the dynamic relaxation from the groundstate. Other
critical exponents are also calculated, as shown in Table.~\ref{t1}.
Except for the transition temperature, the critical exponents in
both Case-I and Case-III are very close to the ones of the $2D$
Ising model, showing that they are both in the Ising universality.

Now we argue that the agreement of critical exponents for Case-I,
Case-III and the $2D$ Ising model is not an accident. The present
model, though with vacancies, has a perfect symmetry that each site
has three equivalent neighbors preserving both
$\sqrt{5}\times\sqrt{5}$ translational and four-fold rotational
invariances of the lattice structure \cite{che11}. In particular,
Case-III is invariant under a {\it block-spin rotation}: $S_i
\rightarrow S_i e^{in\pi}$, associated with a mapping
$J_{1}\rightarrow J_{1},\,J_{1}'\rightarrow - J_{1}'$. Here, the
integer $n$ denotes the block index, and $J_1$ ($J_1'$) indicates
the coupling of intrablock (interblock). Therefore, the model is
equivalent to the ferromagnetic Ising model defined on the square
lattice with the vacancy superstructure. Then, a topological
deformation can be performed from the square lattice with the
$\sqrt{5} \times \sqrt{5}$ vacancy order to the bathroom-tile
lattice, as illustrated in Fig.~\ref{f6}(b). Hence Case-III and the
$2D$ bathroom-tile ferromagnetic Ising model are equivalent.
Remarkably, the latter model (with the NN coupling only) is exactly
solvable, with the exact inverse Curie temperature $K_c=tanh^{-1}
(\sqrt{(5+4\sqrt{2})/2} -1-1/\sqrt{2})\approx 0.6951$ \cite{cod10}.
This value is in perfect agreement with our numerical value
$0.6952(1)$ of Case-III. In addition, it is known that
the Ising models defined on the square, triangular, Kagome, and
bathroom-tile lattices belong to the same universality class
\cite{zhe98,luo09,loh08,bax88,mal12}. As a consequence, identical
critical exponents are predicted between Case-III and the $2D$ Ising
model, as revealed in our numerical results. We note that the
critical temperature $T_c$ is lower in Case-III than in the $2D$
Ising mode. This is clearly due to the existence of
vacancies (with the concentration $20\%$, corresponding to $x=0.4$
in realistic A$_{y}$Fe$_{2-x}$Se$_{2}$ compounds), which in turn
leads to three NN bonds for each iron site.

Similar analysis can also be carried out for Case-I and Case-II.
Under the block-spin rotation and the topological transformation,
the bathroom-tile Ising model with both the ferromagnetic NN and NNN
couplings is derived from Case-I. Since the
ferromagnetic NNN coupling is irrelevant to the Ising universality,
the critical exponents agree well with those of the $2D$ Ising
model. However, it significantly enhances the
transition temperature $T_c$ due to the increase of the
ferromagnetic coupled bonds. By contrast, the situation is quite
different in Case-II, because the magnetic frustration between the
antiferromagnetic NN and NNN couplings exists still even after the
mapping. It explains why the individual exponents are non-universal
(may vary with model parameters in the same SAFM phase) and
different to those of the $2D$ Ising model. Meanwhile, the
transition temperature $T_c$ is also suppressed.

   We note that for the parameters obtained from the first-principles
calculations, i.e., $J_1 = -43$ meV, $J_2 = -4.5$ meV,
$J_1'=-14.5$ meV, and $J_2'=19$ meV \cite{cao11}, above mapping
leads to dominating ferromagnetic couplings and small
antiferromagnetic couplings. Therefore, the magnetic frustration is
actually suppressed in realistic A$_{y}$Fe$_{2-x}$Se$_2$ compounds,
indicating that they are in the same BSC phase as Case-I.
In order to directly compare with experiments,
additional simulations are performed with these model
parameters. As expected, the transition temperature $T_N = 545 K$ is
determined, compatible with the experimental result $T_N \sim 550K$,
and all of the critical exponents, such as $\beta = 0.121(2), \nu =
0.98(1), z = 2.16(1), \theta = 0.189(2), \gamma = 1.69(2)$, and
$\lambda = 1.59(1)$, are in good agreement with those in Case-I.

\section{Conclusion}
\begin{figure}[ht]
\epsfysize=6.5cm \epsfclipoff \fboxsep=0pt
\setlength{\unitlength}{1.cm}
\begin{picture}(10,6)(0,0)
\put(-3.0,-0.3){{\epsffile{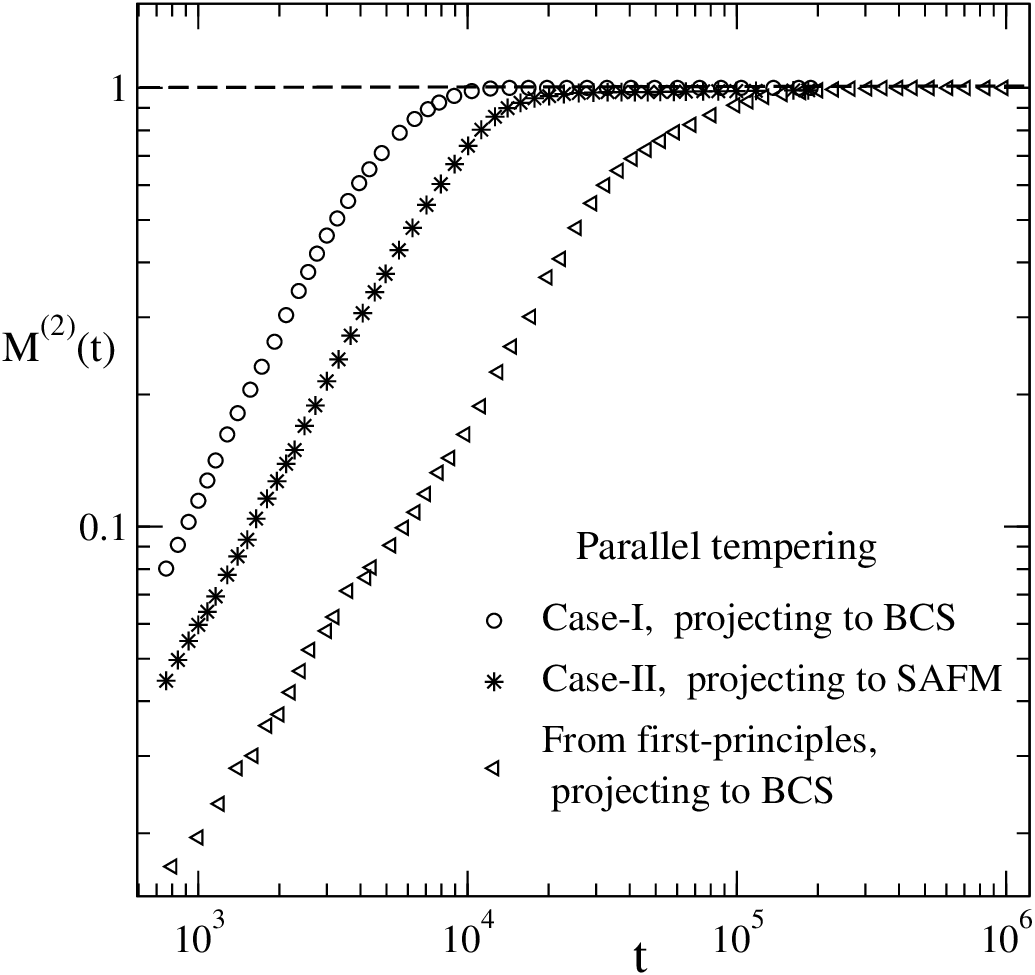}}}\epsfysize=6.5cm
\put(5.2,-0.3){{\epsffile{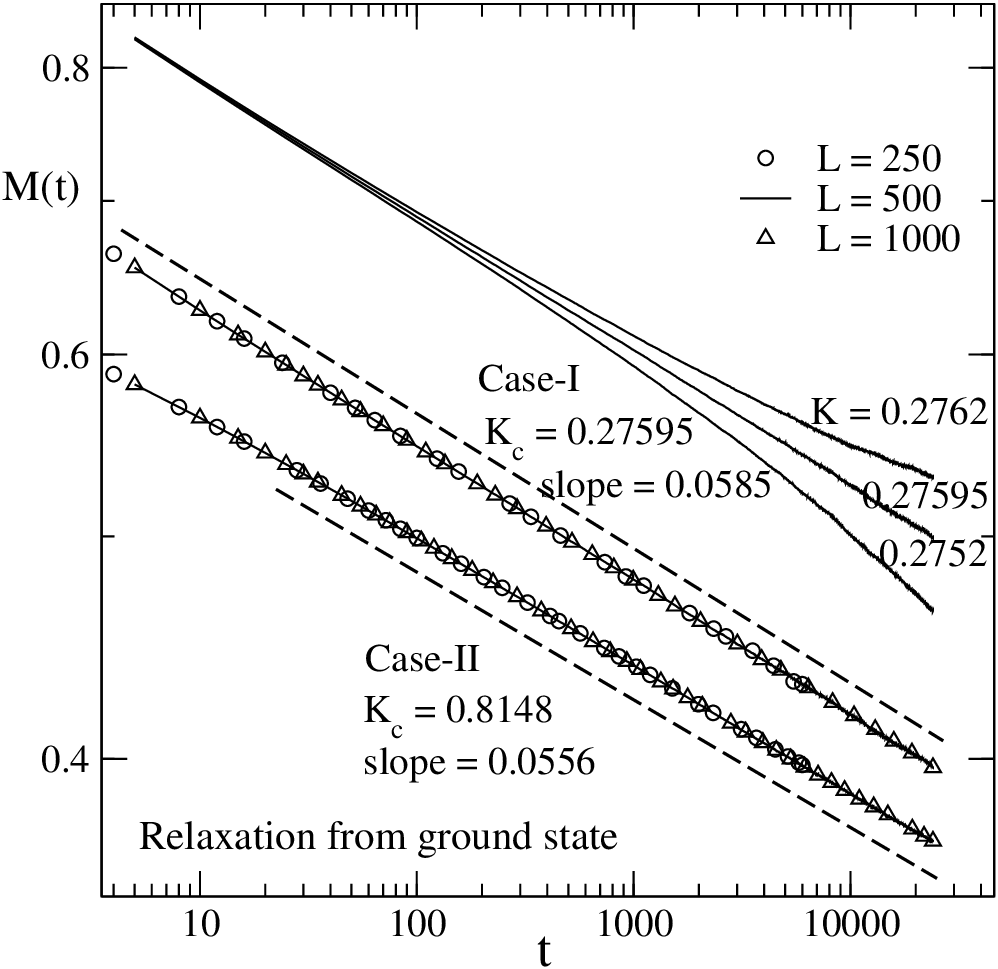}}}
\end{picture}

\hspace{1.0cm}\footnotesize{(a)}\hspace{8.0cm}\footnotesize{(b)}
\caption{(a) The second moment of the pseudo-magnetization in the
parallel tempering process is displayed for different sets of model
parameters. Dashed line represents constant, $M^{(2)}(t) = 1$.
\quad(b) Dynamic relaxation of the pseudo-magnetization is plotted
for different temperatures. For clarity, the curves at $K_c$ with
different lattice sizes are shifted down. Dashed lines indicate
power-law fits.}\label{f2}
\end{figure}

Using the parallel tempering technique and short-time dynamic
approach, we have numerically investigated the finite-temperature
magnetic phase transitions in the extended $J_1$-$J_2$ Ising spin
lattice with the $\sqrt{5} \times \sqrt{5}$ vacancy superstructure,
for Case-I, Case-II, and Case-III, defined in
Eq.~(\ref{f2}). The vacancy concentration $p = 20\%$ is considered
in the simulations, and results including the transition temperature
and critical exponents are summarized in Table~\ref{t1}.

(i) The magnetic configuration of the groundstate in Case-I is
{ detected to be} the BSC state as observed in realistic
A$_{y}$Fe$_{2-x}$Se$_2$ compounds with $x \sim 0.4$.
While the groundstate magnetic configuration in Case-II is the SAFM
state as observed in other iron pnictides without iron
vacancies.

(ii) A dramatic enhancement in the transition temperature,
$T_c=3.6238(4)$, is determined in Case-I. This value is almost three
times as large as that in Case-II where $T_c=1.2273(1)$. It is quite
compatible with the corresponding magnetic transition temperatures
of the BSC ($T_N\sim 550$ K) \cite{bao11,ye11} and SAFM ($T_N\sim
100-200$ K) \cite{cru08,che09} phases reported respectively in
experiments.

(iii) The nature of the magnetic transition in Case-I is revealed
of the $2D$ Ising universality class, while the
deviation of the critical exponents from those in the $2D$ Ising
model, reaching about $10\%$, indicates that Case-II belongs to the
Suzuki's weak universality class.

(iv) Case-III is shown to be equivalent to the bathroom-tile Ising
model which is exactly solvable. Good agreement between the
numerical results and the exact solution demonstrates the validity
of our numerical simulations on this class of complex systems.

Basing on these numerical results, we conclude that
the dramatic enhancement of the transition temperature in the BSC
state as observed in realistic materials A$_{y}$Fe$_{2-x}$Se$_2$
should be mainly due to a combination effect of the perfect vacancy
superstructure and the block lattice contraction. The latter in turn
leads to the suppression of magnetic frustrations due to the
ferromagnetic intrablock couplings and the dominating
antiferromagnetic interblock NNN coupling. It is also
supported by the numerical simulations with the model parameters
obtained from the first-principles calculations, where the
transition temperature $T_N= 545$ K is compatible with the
experimental results $T_N \sim 550$ K.

{\bf Acknowledgements:} We would like to thank helpful discussions
with C. Cao. This work was supported in part by the National Natural
Science Foundation of China (under Grant Nos. 11205043, 11075137 and
11274084).

\bibliography{zheng,super}

\begin{thebibliography}{10}

\bibitem{kam08}
{Y. Kamihara, T. Watanabe, M. Hirano, and H. Hosono}, J. Am. Chem. Soc. {\bf
  130},  3296  (2008).

\bibitem{che08}
{X.H. Chen, T. Wu, R.H. Liu, H. Chen, and D.F. Fang}, Nature {\bf 453},  761
  (2008).

\bibitem{che08b}
{G.F. Chen, Z. Li, D. Wu, G. Li, W.Z. Hu, J. Dong, P. Zheng, J.L. Luo, and N.L.
  Wang}, Phys. Rev. Lett. {\bf 100},  247002  (2008).

\bibitem{ren08b}
{Z.A. Ren, W. Lu, J. Yang, W. Yi, X.L. Shen, C. Zheng, G.C. Che, X.L. Dong,
  L.L. Sun, F. Zhou, and Z.X. Zhao}, Chin. Phys. Lett. {\bf 25},  2215  (2008).

\bibitem{wan08}
{C. Wang, L.J. Li, S. Chi, Z.W. Zhu, Z. Ren, Y.K. Li, Y.T. Wang, X. Lin, Y.K.
  Luo, S. Jiang, X.F. Xu, G.H. Cao, and Z.A. Xu}, Europhys. Lett. {\bf 83},
  67006  (2008).

\bibitem{pag10}
{J. Paglione and R.L. Greene}, Nature Phys. {\bf 6},  645  (2010).

\bibitem{rot08}
{M. Rotter, M. Tegel, and D. Johrendt}, Phys. Rev. Lett. {\bf 101},  107006
  (2008).

\bibitem{wan08b}
{X.C. Wang, Q.Q. Liu, Y.X. Lv, W.B. Gao, L.X. Yang, R.C. Yu, F.Y. Li, and C.Q.
  Jin}, Solid State Commun. {\bf 148},  538  (2008).

\bibitem{hsu08}
{F.C. Hsu, J.Y. Luo, K.W. Yeh, T.K. Chen, T.W. Huang, P.M. Wu, Y.C. Lee, Y.L.
  Huang, Y.Y. Chu, D.C. Yan, and M.K. Wu }, Proc. Natl. Acad. Sci. U.S.A. {\bf
  105},  14262  (2008).

\bibitem{cru08}
{C. de la Cruz, Q. Huang, J.W. Lynn, J.Y. Li, W. Ratcliff II, J.L. Zarestky,
  H.A. Mook, G.F. Chen, J.L. Luo, N.L. Wang, and P.C. Dai }, Nature {\bf 453},
  899  (2008).

\bibitem{che09}
{H. Chen, Y. Ren, Y. Qiu, W. Bao, R.H. Liu, G. Wu, T. Wu, Y.L. Xie, X.F. Wang,
  Q. Huang, and X.H. Chen}, Europhys. Lett. {\bf 85},  17006  (2009).

\bibitem{bao09}
{W. Bao, Y. Qiu, Q. Huang, M.A. Green, P. Zajdel, M.R. Fitzsimmons,
  M.Zhernenkov, S. Chang, M.H. Fang, B. Qian, E.K. Vehstedt, J.H. Yang, H.M.
  Pham, L. Spinu, and Z.Q. Mao}, Phys. Rev. Lett. {\bf 102},  247001  (2009).

\bibitem{sin09}
{D.J. Singh}, Physica C: Superconductivity {\bf 469},  418  (2009).

\bibitem{joh10}
{D.C. Johnston}, Advances in Physics {\bf 59},  803  (2010).

\bibitem{maz11}
{I. Mazin}, Physics {\bf 4},  26  (2011).

\bibitem{sin08}
{D.J. Singh and M.H. Du}, Phys. Rev. Lett. {\bf 100},  237003  (2008).

\bibitem{cao08}
{C. Cao, P.J. Hirschfeld, and H.P. Cheng}, Phys. Rev. {\bf B 77},  220506(R)
  (2008).

\bibitem{yil08}
{T. Yildirim}, Phys. Rev. Lett. {\bf 101},  057010  (2008).

\bibitem{ma08}
{F.J. Ma, Z.Y. Lu, and T. Xiang}, Phys. Rev. {\bf B 78},  224517  (2008).

\bibitem{si08}
{Q.M. Si and E. Abrahams}, Phys. Rev. Lett. {\bf 101},  076401  (2008).

\bibitem{guo10}
{J.G. Guo, S.F. Jin, G. Wang, S.C. Wang, K.X. Zhu, T.T. Zhou, M. He, and X.L.
  Chen}, Phys. Rev. {\bf B 82},  180520(R)  (2010).

\bibitem{fan11}
{M.H. Fang, H.D. Wang, C.H. Dong, Z.J. Li, C.M. Feng, J. Chen, and H.Q. Yuan},
  Europhys. Lett. {\bf 94},  27009  (2011).

\bibitem{wan11}
{H.D. Wang, C.H. Dong, Z.J. Li, Q.H. Mao, S.S. Zhu, C.M. Feng, H.Q. Yuan, and
  M.H. Fang}, Europhys. Lett. {\bf 93},  47004  (2011).

\bibitem{bao11}
{W. Bao, Q.Z. Huang, G.F. Chen, M.A. Green, D.M. Wang, J.B. He, and Y.M. Qiu},
  Chin. Phys. Lett. {\bf 28},  086104  (2011).

\bibitem{ye11}
{F. Ye, S. Chi, W. Bao, X.F. Wang, J.J. Ying, X.H. Chen, H.D. Wang, C.H. Dong,
  and M.H. Fang}, Phys. Rev. Lett. {\bf 107},  137003  (2011).

\bibitem{pom11}
{V.Y. Pomjakushin, E.V. Pomjakushina, A.K. Maziopa, K. Conder, and Z.
  Shermadini}, J. Phys. Cond. Mat. {\bf 23},  156003  (2011).

\bibitem{wan11b}
{Z. Wang, Y.J. Song, H.L. Shi, Z.W. Wang, Z. Chen, H.F. Tian, G.F. Chen, J.G.
  Guo, H.X. Yang, and J.Q. Li}, Phys. Rev. {\bf B 83},  140505(R)  (2011).

\bibitem{cao11}
{C. Cao and J.H. Dai}, Phys. Rev. Lett. {\bf 107},  056401  (2011).

\bibitem{yan11}
{X.W. Yan, M. Gao, Z.Y. Lu, and T. Xiang}, Phys. Rev. {\bf B 83},  233205
  (2011).

\bibitem{yu11}
{R. Yu, P. Goswami, and Q.M. Si}, Phys. Rev. {\bf B 84},  094451  (2011).

\bibitem{fan12}
{C. Fang, B. Xu, P.C. Dai, T. Xiang, and J.P. Hu}, Phys. Rev. {\bf B 85},
  134406  (2012).

\bibitem{hu12}
{J.P. Hu, B. Xu, W.M. Liu, N.N. Hao, and Y.P. Wang}, Phys. Rev. {\bf B 85},
  144403  (2012).

\bibitem{liu11}
{R.H. Liu, X.G. Luo, M. Zhang, A.F. Wang, J.J. Ying, X.F. Wang, Y.J. Yan, Z.J.
  Xiang, P. Cheng, G.J. Ye, Z.Y. Li, and X.H. Chen}, Europhys. Lett. {\bf 94},
  27008  (2011).

\bibitem{che11b}
{F. Chen, M. Xu, Q.Q. Ge, Y. Zhang, Z.R. Ye, L.X. Yang, J. Jiang, B.P. Xie,
  R.C. Che, M. Zhang, A.F. Wang, X.H. Chen, D.W. Shen, J.P. Hu, and D.L. Feng},
  Phys. Rev. {\bf X 1},  021020  (2011).

\bibitem{cao11c}
{C. Cao and J.H. Dai}, Phys. Rev. {\bf B 83},  193104  (2011).

\bibitem{yan11b}
{X.W. Yan, M. Gao, Z.Y. Lu, and T. Xiang}, Phys. Rev. Lett. {\bf 106},  087005
  (2011).

\bibitem{yu11b}
{R. Yu, J.X. Zhu, and Q. Si}, Phys. Rev. Lett. {\bf 106},  186401  (2011).

\bibitem{zho11}
{Y. Zhou, D.H. Xu, F.C. Zhang, and W.Q. Chen}, Europhys. Lett. {\bf 95},  17003
   (2011).

\bibitem{wan11c}
{F. Wang, F. Yang, M. Gao, Z.Y. Lu, T. Xiang, and D.H. Lee}, Europhys. Lett.
  {\bf 93},  57003  (2011).

\bibitem{zha11}
{G.M. Zhang, Z.Y. Lu, and T. Xiang}, Phys. Rev. {\bf B 84},  052502  (2011).

\bibitem{che11}
{H. Chen, C. Cao, and J.H. Dai}, Phys. Rev. {\bf B 83},  180413(R)  (2011).

\bibitem{jia12}
{H.M. Jiang, W.Q. Chen, Z.J. Yao, and F.C. Zhang}, Phys. Rev. {\bf B 85},
  104506  (2012).

\bibitem{sca12}
{D.J. Scalapino}, arXiv:1207.4093,  to appear in Rev. Mod. Phys.  (2012).

\bibitem{wan11d}
{ M.Y. Wang, C. Fang, D.X. Yao, G.T. Tan, L.W. Harriger, Y. Song, T. Netherton,
  C.L. Zhang, M. Wang, M.B. Stone, W. Tian, J.P. Hu, and P.C. Dai}, Nat.
  Commun. {\bf 2},  580  (2011).

\bibitem{yan11c}
{X.W. Yan, M. Gao, Z.Y. Lu, and T. Xiang}, Phys. Rev. {\bf B 84},  054502
  (2011).

\bibitem{cao11b}
{C. Cao and J.H. Dai}, Chin. Phys. Lett. {\bf 28},  057402  (2011).

\bibitem{ken08}
{R. Kenna and J.J. Ruiz-Lorenzo}, Phys. Rev. {\bf E 78},  031134  (2008).

\bibitem{rom09}
{F. Rom\'a, S.R. Gusman, A.J.R. Pastor, F. Nieto, and E.E. Vogel}, Physica {\bf
  A 388},  2821  (2009).

\bibitem{yin09}
{J.Q. Yin and D.P. Landau}, Phys. Rev. {\bf E 80},  051117  (2009).

\bibitem{jin12}
{S.B. Jin, A. Sen, and A.W. Sandvik}, Phys. Rev. Lett. {\bf 108},  045702
  (2012).

\bibitem{zhe98}
B. Zheng, Int. J. Mod. Phys. {\bf B12},  1419  (1998).

\bibitem{luo98}
{H.J. Luo, L. Sch\"ulke, and B. Zheng}, Phys. Rev. Lett. {\bf {81}},  180
  (1998).

\bibitem{oze07}
{Y. Ozeki and N. Ito}, J. Phys. A: Math. Theor. {\bf 40},  R149  (2007).

\bibitem{alb11}
{E.V. Albano, M.A. Bab, G. Baglietto, R.A. Borzi, T.S. Grigera, E.S. Loscar,
  D.E. Rodr{\'i}guez, M.L.R. Puzzo, and G.P. Saracco}, Rep. Prog. Phys. {\bf
  74},  026501  (2011).

\bibitem{gra05}
{E. Granato and D. Dom{\'i}nguez}, Phys. Rev {\bf B 71},  094521  (2005).

\bibitem{cal05}
{P. Calabrese and A. Gambassi}, J. Phys. A {\bf 38},  R133  (2005).

\bibitem{lei07}
{X.W. Lei and B. Zheng}, Phys. Rev. {\bf E75},  040104  (2007).

\bibitem{zho09}
{N.J. Zhou, B. Zheng, and Y.Y. He}, Phys. Rev. {\bf B80},  134425  (2009).

\bibitem{zho10}
{N.J. Zhou, B. Zheng, and D.P. Landau}, Europhys. Lett. {\bf 92},  36001
  (2010).

\bibitem{zho10a}
{N.J. Zhou and B. Zheng}, Phys. Rev. {\bf E82},  031139  (2010).

\bibitem{qin12}
{X.P. Qin, B. Zheng, and N.J. Zhou}, J. Phys. A: Math. Theor. {\bf 45},  115001
   (2012).

\bibitem{jan89}
{H.K. Janssen, B. Schaub, and B. Schmittmann}, Z. Phys. {\bf B 73},  539
  (1989).

\bibitem{tan87}
{S. Tang and D.P. Landau}, Phys. Rev. {\bf B 36},  567  (1987).

\bibitem{cod10}
{A. Codello}, J. Phys. A: Math. Theor. {\bf 43},  385002  (2010).

\bibitem{luo09}
{Z.H. Luo, L. Mushtaq, Y. Liu, and J.R. Lin}, Chin. Phys. {\bf B 18},  2696
  (2009).

\bibitem{loh08}
{Y.L. Loh, D.X. Yao, and E.W. Carlson}, Phys. Rev. {\bf B 77},  134402  (2008).

\bibitem{bax88}
{R.J. Baxter and T.C. Choy}, J. Phys. A: Math. Gen. {\bf 21},  2143  (1988).

\bibitem{mal12}
{A. Malakis, G. Gulpinar, Y. Karaaslan, T. Papakonstantinou, and G. Aslan},
  Phys. Rev. {\bf E 85},  031146  (2012).

\end{thebibliography}
\bibliographystyle{prsty}

\begin{table}[h]\centering
\caption{The inverse transition temperatures and critical exponents
obtained with the short-time dynamic approach are listed for Case-I,
Case-II, and Case-III, in comparison with those of $2D$ Ising model
on the square lattice from literatures \cite{zhe98,alb11,lei07}. Not
all of the critical exponents are independent, and the scaling laws
$\gamma/\nu + 2\beta/\nu =d$ and $\lambda + \theta z = d $ hold
quite well within error bars in each case.}
\begin{tabular}[t]{l p{10mm} l l l l}
\hline
\hline           &            & Case-I            & Case-II       & Case-III     & $2D$ Ising \\
\hline
Ground-state     & $K_c$      &  $0.27595(3)$     & $0.8148(1)$   & $0.6952(1)$  &  $0.44069$   \\
                 & $\beta$    &  $0.125(2)$       & $0.109(1)$    & $0.122(2)$   &  $1/8$       \\
                 & $\nu$      &  $1.00(1)$        & $0.90(1)$     & $1.00(2)$    &  $1$         \\
                 & $z$        &  $2.16(1)$        & $2.17(1)$     & $2.18(2)$    &  $2.16(1)$\\
                 & $\beta/\nu$ & $0.125(2)$       & $0.121(2)$    & $0.122(3)$   &  $1/8$    \\  \hline
Disordered       & $\theta$   &  $0.186(2)$       & $0.167(1)$    & $0.189(1)$   &  $0.191(1)$ \\
                 & $\gamma $  &  $1.72(2)$        & $1.58(2)$     & $1.76(3)$    &  $7/4$    \\
                 & $\gamma/\nu$& $1.74(2)$        & $1.75(2)$     & $1.77(2)$    &  $7/4$    \\
                 & $\lambda$  &  $1.59(1)$        & $1.65(1)$     & $1.60(1)$    &  $1.59(1)$ \\
\hline \hline
\end{tabular}
\label{t1}
\end{table}

\clearpage

\end{document}